\documentclass[12pt,preprint]{aastex}

\newcommand{\eg}{{\it e.g.}}
\newcommand{\etal}{{\it et al.}}

\begin{document}

\title{The Size-Frequency Distribution of \\
Dormant Jupiter Family Comets}

\author{Kathryn Whitman$^1$, Alessandro Morbidelli$^2$ and Robert Jedicke$^1$}
\affil{$^1$Institute for Astronomy, University of Hawaii, 2680 Woodlawn Dr., Honolulu, HI 96822}
\affil{$^2$Observatory of Nice, B.P. 4229, 06304 Nice Cedex 4}

\begin{abstract}
We estimate the total number and the slope of the size frequency
distribution (SFD) of dormant Jupiter Family Comets (JFCs) by fitting
a one-parameter model to the known population.  We first select 61
Near Earth Objects (NEOs) that are likely to be dormant JFCs because
their orbits are dynamically coupled to Jupiter \citep{Bot02}. Then,
from the numerical simulations of \citet{Lev97}, we construct an orbit
distribution model for JFCs in the NEO orbital element space. We
assume an orbit independent SFD for all JFCs, the slope of which is
our unique free parameter.  Finally, we compute observational biases
for dormant JFCs using a calibrated NEO survey simulator
\citep{Jed03}. By fitting the biased model to the data, we estimate
that there are $\sim 75$ dormant JFCs with $H<18$ in the NEO region
and that the slope of their cumulative SFD is $-1.5 \pm 0.3$. Our
slope for the SFD of dormant JFCs is very close to that of active JFCs
as determined by \citet{Wei03}.  Thus, we argue that when JFCs fade
they are likely to become dormant rather than to disrupt and that the
fate of faded comets is size independent.  Our results imply that the
size distribution of the JFC progenitors --the scattered disk
trans-Neptunian population-- either (i) has a similar and shallow SFD
or (i$^\prime$) is slightly steeper and physical processes acting on
the comets in a size-dependent manner creates the shallower active
comet SFD.  Our measured slope, typical of collisionally evolved
populations with a size dependent impact strength \citep{Ben99},
suggests that scattered disk bodies reached collisional equilibrium
inside the proto-planetary disk prior to their removal from the
planetary region.
\end{abstract}


\section{Introduction}

The populations of small bodies in the solar system provide an
important key to unlocking its birth and evolution.  The constituent
materials of comets can be explored through spectroscopic
observations, while their dynamical and physical history can be
extracted from studies of their orbital and size-frequency
distribution.

Several studies have traced the dynamical evolution of comets from
their parent reservoirs (the scattered disk for Jupiter family comets,
the Oort cloud for long period comets) to their ultimate ejection from
the Solar system \citep{Wei78,Fer81,Lev97,Wie99}. The orbital
distribution of comets that these studies predict is different from
the observed distribution. In general, there is a deficit of comets on
dynamically evolved orbits, which is usually interpreted as an
indication that comets {\it fade} with age: comets appear to be active
only over a limited number of revolutions with small perihelion
distance.

Curiously, Jupiter family comets (JFC) and long period comets (LPC)
appear to follow very different fading laws. The JFCs are active over
a lifetime of about 10,000~y or $\sim 1,000$ revolutions \citep{Lev97}
while the LPCs disappear much faster. Only 10\% of the LPCs survive
more than 50 passages to small perihelion, while only 1\% of them
survives more than 2000 passages \citep{Wie99}.

The fate of faded comets is still a subject of debate. Do comets
disintegrate into small, undetectable pieces (like in the case of
comet LINEAR C/2001 A2) or do they develop an insulating crust that
prevents or restricts further outgassing allowing them to survive as
inactive (dormant or extinct) bodies with an asteroidal appearance?
In this respect, JFCs and LPCs seem to behave differently.
\citet{Bot00} and \citet{Bot02} concluded that about $6 \pm 4$\% of
the population of Near Earth Objects (NEOs) is composed of extinct
JFCs. Recent spectral observations of NEOs \citep{Fer01,Bin04}
corroborate this result.  This implies that a substantial fraction of
JFCs orbit the Sun as inactive bodies.  Conversely, \citet{Lev02}
showed that the population of dormant LPCs is only about 1\% of the
total expected population if it is assumed that LPCs fade rather than
disrupt.  In other words, 99\% of LPCs disintegrate when they
disappear.

The papers by Bottke \etal\ (2000,2002) were not devoted to
characterizing the population of extinct JFCs.  The authors created an
orbital distribution model for the overall NEO population by combining
the characteristic distributions of objects coming from 5 possible
source regions: 4 sources belonging to the main asteroid belt, plus
the JFCs. Each source population was assumed to have the same absolute
magnitude ($H$) distribution.  The relative importance of the sources,
as well as the slope of the $H$-distribution, were determined by the
best fit to the distribution of the NEOs detected by the Spacewatch
survey (138 objects). Their results concerning JFCs were hampered by
small number statistics since the Spacewatch sample included only a
handful of dormant comet candidates.  In particular, the NEO
$H$-distribution determined by Bottke \etal\ (2000,2002) does not
necessarily characterize the dormant JFC population.

The size frequency distribution (SFD) of dormant comets is important
to understanding the fading issue.  It can be related to the $H$
frequency distribution (HFD) and to the mass distribution (MFD)
through a simple calculation if one assumes that the objects have a
size-independent albedo and density \citep{Dur93}.  Since other
researchers have determined the SFD or MFD we provide the conversion
for the reader's convenience.  A cumulative $H$ distribution of the form
\begin{equation}
N(<H)\propto 10^{\alpha H}
\label{e.HFD}
\end{equation}
is strictly equivalent to a cumulative size distribution of the form
\begin{equation}
N(>D)\propto D^{-5\alpha}\ ,
\label{e.SFD}
\end{equation}
where $D$ is the diameter, or to a cumulative mass distribution of the form
\begin{equation}
N(>M)\propto M^{-5\alpha/3}\ ,
\label{e.MFD}
\end{equation}
where $M$ is the mass.  Note that the exponent of the differential and
cumulative HFD are identical while the exponent for the differential
distributions in size and mass become $-5\alpha-1$ and $-5\alpha/3-1$
respectively.  We prefer to work with the $H$-distribution because it
is directly related to the observations (the apparent brightness of an
object is usually measured, not its size or its mass).

In a reservoir of small bodies in collisional equilibrium, like the
asteroid belt or the Kuiper belt, and if the strength of objects is
independent of their size \citep[a Self-Similar Collision Cascade;
\eg][]{Tan96}, then the population is expected to have a SFD with
$\alpha=0.5$ \citep[\eg][]{Doh69}.  In reality, the SFD or
$H$-distribution has a wavy aspect such that the value of $\alpha$ for
the main belt (MB) asteroids is size-dependent with a value that jumps
around the theoretically expected value
\citep[\eg][]{Ive01,Jed98,van70,Kui58}.  There are a variety of
explanations for the variation of $\alpha$ about the theoretical
value, including it being 
\begin{list}{}{}
\item[a)] a relic of the primordial SFD for the larger asteroids
      \citep[\eg][]{Bot04},
\item[b)] a consequence of a size-dependent strength of asteroids
      \citep[\eg][]{Dur98},
\item[c)] due to the quick removal of very small dust particles that
      eliminates the tiny tail of the population that wags the rest of the
      SFD \citep[\eg][]{Cam94}.
\end{list}
The actual SFD is probably a result of a combination of all these
proposed mechanisms and an accurate measure of the SFD of MB asteroids
over all size ranges may allow the effects to be disentangled.

The Near Earth Objects (NEOs) are expected to mimic the wavy SFD of their
source region, the main belt, but to have a larger slope because the
main process that tranports main belt objects into the NEO region --
the Yarkovsky effect \citep{Bot02b} -- is size dependent. This has
been verified for NEOs with $H<18$ by \citet{Mor03}.

For active comets, the SFD may be further complicated by a
size-dependent fading law. Moreover, the comparison between the SFD of
the dormant population with that of the active population can tell us
whether the probability of disintegrating versus becoming dormant is
size dependent.  The comparison between the SFDs of dormant comet
populations that have very different fading laws can also be very
instructive.  Unlike NEOs, comets originate in reservoirs that might
not be in collisional equilibrium. Indeed comets, once stored in the
scattered disk or in the Oort cloud, might have avoided collisions due
to the huge volume available in these reservoirs.

In this paper, taking advantage of the fact that several dozens of
NEOs have been recently discovered on orbits that are typical of JFCs,
we wish to reassess the issue of the total number of dormant JFCs and
of their SFD. In \S\ref{s.Method} we explain the general principle of
our method.  In particular, we explain how we select the candidate
dormant JFCs from the NEO catalogue
(\S\ref{ss.TheKnownJFCPopulation}), how we construct an orbital
distribution model (\S\ref{ss.ModelforNEOsofJFCProvenance}), how we
estimate the observational biases (\S\ref{s.BiasDetermination}) and
how we determine the best fit value of $\alpha$ (\S\ref{ss.ML}). In
Section \ref{s.ModelResults} we compare our best fit model with the
observed population in terms of orbital and absolute magnitude
distributions.  We evaluate the statistical agreement between model
and observations (\S\ref{ss.QuantifyingTheStatisticalAgreement}), and
we discuss the dependence of the results on the selected candidate JFC
population and the systematic errors in our measurement
(\S\ref{ss.Dependence} and \S\ref{ss.SystematicErrors}
respectively). In Section \ref{s.Discussion} we finally discuss the
interesting implications of our results.

\section{Method}
\label{s.Method}

Any set of observed objects is a convolution of the actual underlying
population and the methods and instrumentation used to detect them.
In the case of dormant JFCs currently on NEO orbits, the observed
population distribution $m_{JFC}$ is given by:
\begin{equation}
m_{JFC}(a,e,i,H)\; da\; de\; di\; dH = B(a,e,i,H) \times M_{JFC}(a,e,i,H)\; da\; de\; 
di\; dH 
\end{equation}
where $B(a,e,i,H)$ represents the observational selection effects
(bias) of the survey(s) contributing to the observed population as a
function of the object's $H$, semimajor axis ($a$), eccentricity ($e$)
and inclination ($i$).  $B(a,e,i,H)$ can be thought of as the
probability that an object with $(a,e,i,H)$ has been detected.
$M_{JFC}(a,e,i,H)\; da\; de\; di\; dH$ represents the actual number of
JFCs with orbit elements and $H$ in the ranges $a\rightarrow a+da$,
$e\rightarrow e+de$, $i\rightarrow i+di$ and $H\rightarrow H+dH$.

We assume that the $H$ distribution of the JFCs is independent of their
orbital distribution so that we can write 
\begin{equation}
M_{JFC}(a,e,i,H) = \tilde{f}_{JFC}(a,e,i) \times C_0 \; 10^{\alpha (H-H_0)}
\label{e.aeiH-dist}
\end{equation}
where $\tilde{f}_{JFC}(a,e,i) \; da \; de \; di$ is the fraction of the JFC
population with orbit elements $(a,e,i) \rightarrow (a+da,e+de,i+di)$
and $C_0= \alpha N_0 \ln(10)$, where $N_0$ is the number of JFCs with
$H<H_0$.

Our goal is to determine the slope ($\alpha$) of the $H$-distribution
of dormant JFCs.  As we will see below
(\S\ref{ss.TheKnownJFCPopulation}), $\alpha$ is the only free
parameter in our model distribution $M_{JFC}$. Thus, we can determine
its value by looking for the best 4-dimensional fit in the
$(a,e,i,H)$-space between the model distribution $B\times M_{JFC}$ and
the observed distribution $m_{JFC}$.

To do so, we first need to select the NEOs that are most likely to be
dormant JFCs in order to define the $m_{JFC}$ distribution
(\S\ref{ss.TheKnownJFCPopulation}), then build the distribution model
$M_{JFC}$ in a consistent way
(\S\ref{ss.ModelforNEOsofJFCProvenance}), and evaluate the bias
function $B$ (\S\ref{s.BiasDetermination}). The fitting procedure is
explained in \S\ref{ss.ML}.

\subsection{Selection of dormant JFC candidates}
\label{ss.TheKnownJFCPopulation}

We started from a list of 2677 known NEOs provided by the
Harvard-Smithsonian's MPC data archive on 2004 March 5.  These NEOs
are {\it not} known to be comets, {\it i.e.} none of them are known to
have ever displayed evidence of cometary activity.  Thus, if some of
them are comet nuclei, they are either dormant, extinct, or outgassing
at an undetectable level.  Their absolute magnitudes therefore
correspond to the brightness of their bare nucleus.

\begin{figure}
\includegraphics[angle=0,scale=0.8]{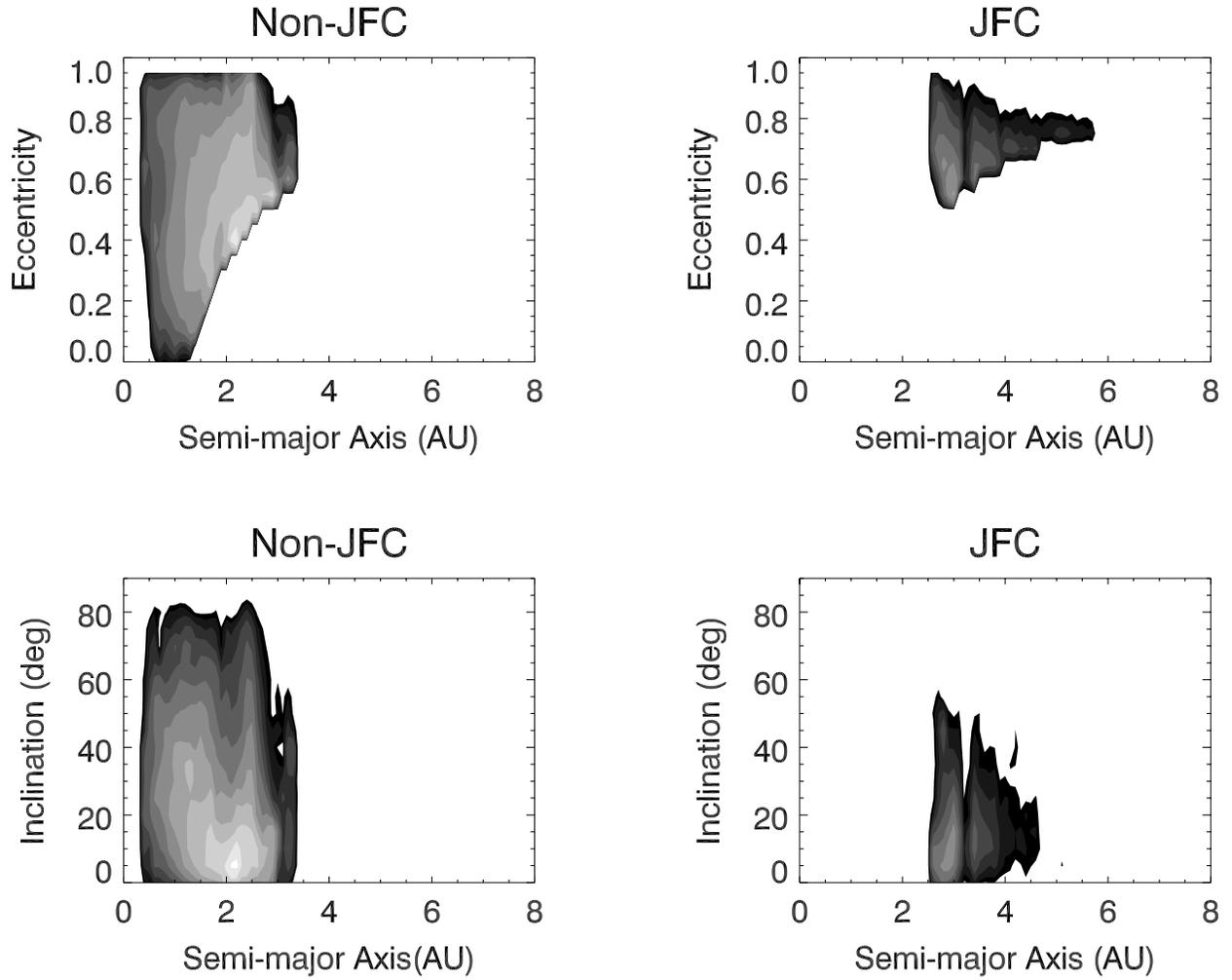}
\caption{(Left)  Orbit distribution for all non-JFC NEOs according to the
\cite{Bot02} model.  (Right)  Orbit distribution for NEOs of JFC
origin according to the same model.}
\label{fig.non-JFC.vs.JFC.aei.distn}
\end{figure}

To identify potential dormant JFCs among the NEOs, we used the
\citet{Bot02} model.  Given a set of values for $a$, $e$ and $i$, the
model provides the probability, $P_{JFC}(a,e,i)$, that the
corresponding object is a JFC.  This is possible because the
characteristic orbital distributions for NEOs of JFC origin and of
asteroidal origin are distinct even though they partially overlap.
Figure \ref{fig.non-JFC.vs.JFC.aei.distn} shows the orbit distribution
for NEOs of non-JFC and JFC provenance according to the \citet{Bot02}
model.  It is important to keep in mind that these figures show the
3-dimensional $(a,e,i)$ space collapsed into 2-dimensions and,
therefore, that the amount of overlap between the orbit distributions
is exaggerated.  The orbit distributions in figure
\ref{fig.non-JFC.vs.JFC.aei.distn} embody all our knowledge of the
source region for the NEOs as a function of their orbit elements.

\begin{figure}[h!]
\includegraphics[angle=0,scale=0.8]{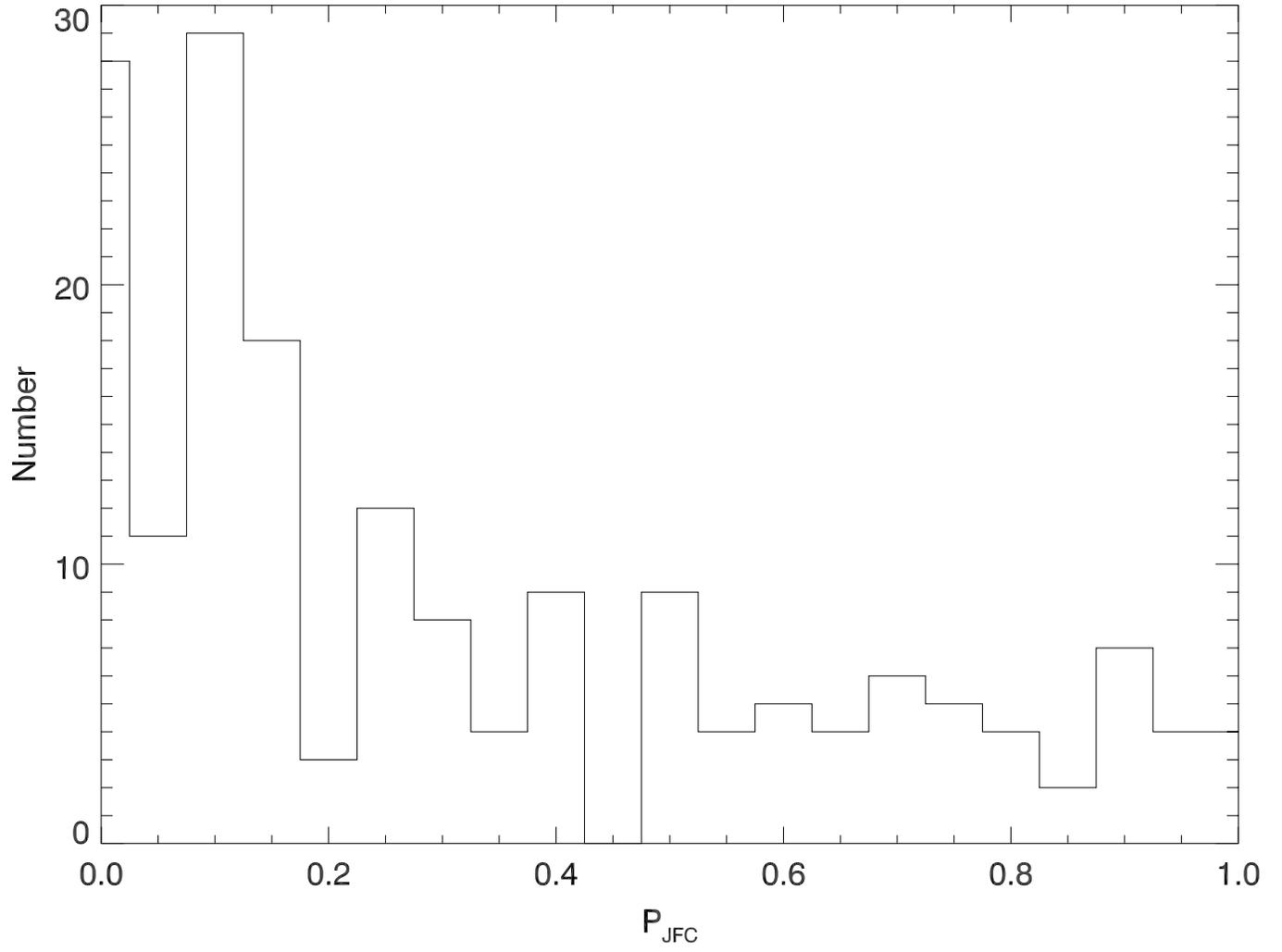}
\caption{Distribution of the probability that known NEOs derive from
  the JFC source region according to the \citet{Bot02} model.  Only
  objects with $P_{JFC}>0.01$ are shown.}
\label{fig.pjfc}
\end{figure}

The condition $P_{JFC}(a,e,i) > 0$ selects 1657 objects.
Nevertheless, for most of these objects $P_{JFC}$ is very small
($<0.01$), so they are likely to be asteroids even if a JFC origin
cannot be ruled out.  Figure \ref{fig.pjfc} shows the distribution of
$P_{JFC}$ for those objects with $P_{JFC}>0.01$ and illustrates that
restricting the NEO sample to objects with increasingly larger
$P_{JFC}$ reduces the fraction of asteroid interlopers. On the other
and, it also reduces the total number of objects and eventually runs
the data into the noise. Thus, there is a trade-off on the choice of
the threshold value of $P_{JFC}$, hereafter denoted by $P_{cutoff}$,
used to select our sample of candidate dormant JFCs.

\begin{figure}[h!]
\includegraphics[angle=0,scale=0.8]{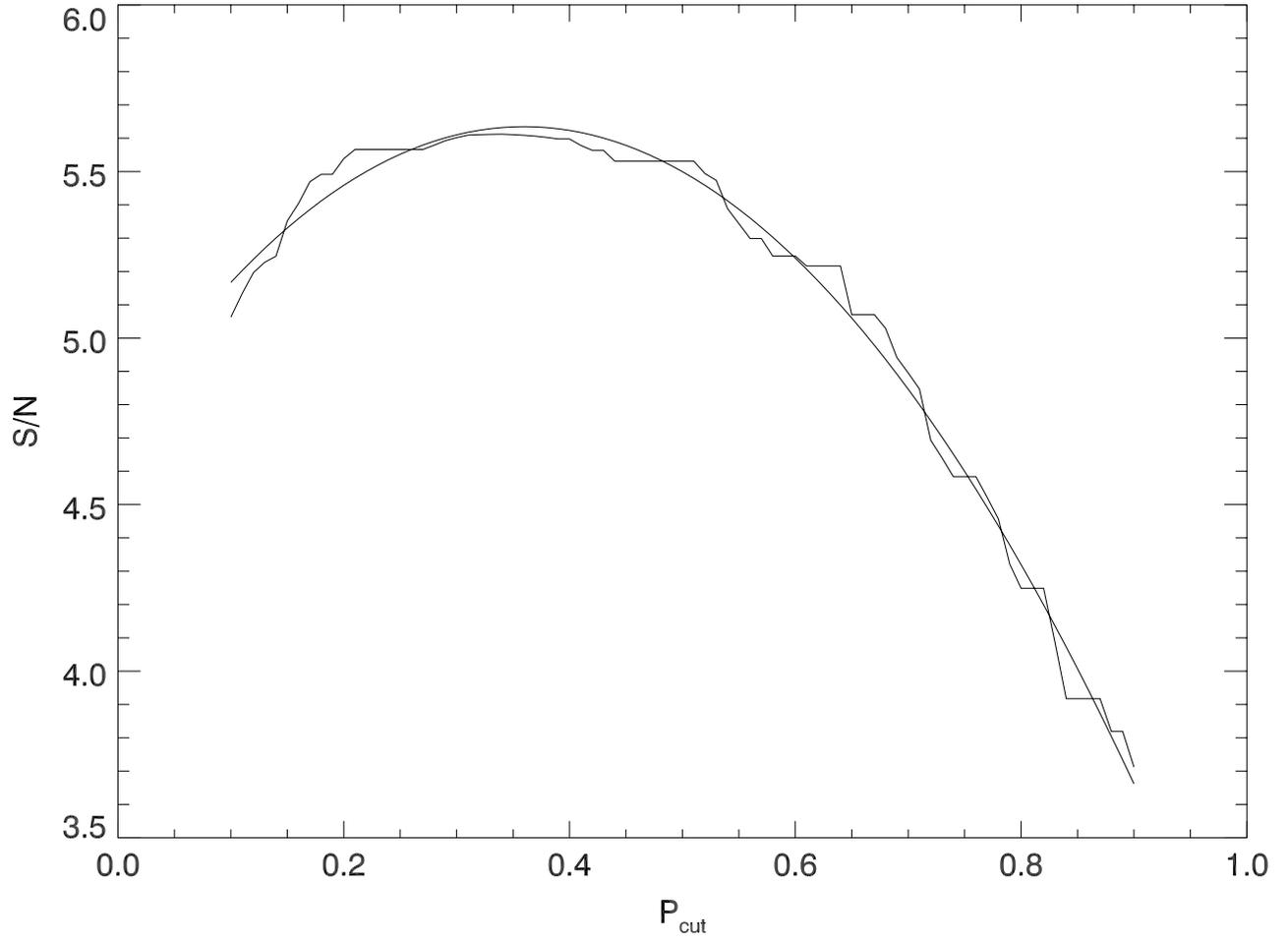}
\caption{The variation of the $S/N$, and the second order polynomial
fit to that curve, as a function of the cutoff on the probability that
the objects are NEOs derived from the JFC source region.  The maximum
$S/N$ occurs at $P_{cutoff} = 0.37$.}
\label{fig.s2b.vs.pjfc}
\end{figure}

We set $P_{cutoff}$ to the value that maximizes the signal-to-noise
ratio carried by the NEOs of JFC origin.  The signal-to-noise ratio
for a specific $P_{cutoff}$ is given by
\begin{eqnarray}
{S \over N}(P_{cutoff}) &=& \frac{Signal}{\sqrt{Signal+Background}} \\
    &=& \frac{\displaystyle\sum_{i}^{P > P_{cutoff}}
  P_{i}}{\sqrt{\displaystyle\sum_{i}^{P > P_{cutoff}}P_{i} + \sum_{i}^{P > P_{cutoff}}(1-P_{i})}} \\
{S \over N}(P_{cutoff}) &=& \frac{\displaystyle\sum_{i}^{P > P_{cutoff}} P_{i}}{\sqrt{\displaystyle\sum_{i}^{P > P_{cutoff}}1}}
\label{e.s2n}
\end{eqnarray}
\noindent where $P_{i}$ is the JFC probability for object $i$. Because
the $S/N$ function is not smooth we fit a second-order polynomial to
the curve. The shape of this function is shown in figure
\ref{fig.s2b.vs.pjfc} where it is clear that a maximum $S/N\sim
5.7$ is achieved at $P_{cutoff} = 0.37$.  There were a total of 67
NEOs with $P_{JFC} \ge 37$\%.  We then confirmed that all but one of
them (2004 CB) had never displayed any cometary activity
(http://www.ifa.hawaii.edu/$\sim$yan/cometlist.html).  That object was
removed from the data set leaving us with 66 candidate dormant JFC
nuclei.

The bias calculation (\S\ref{s.BiasDetermination}) proved to be
computationally intensive so we chose to eliminate five of the JFC
candidates with outlying orbit elements.  This allowed us to restrict
the range of $(a,e,i,H)$ over which we needed to determine the bias
to: $2.6AU \leq a < 3.8AU$, $.55 \leq e < .95$, $0^{\circ} \leq i <
55^{\circ}$, $14. \leq H < 22.5$.  

\begin{figure}
\includegraphics[angle=0,scale=0.8]{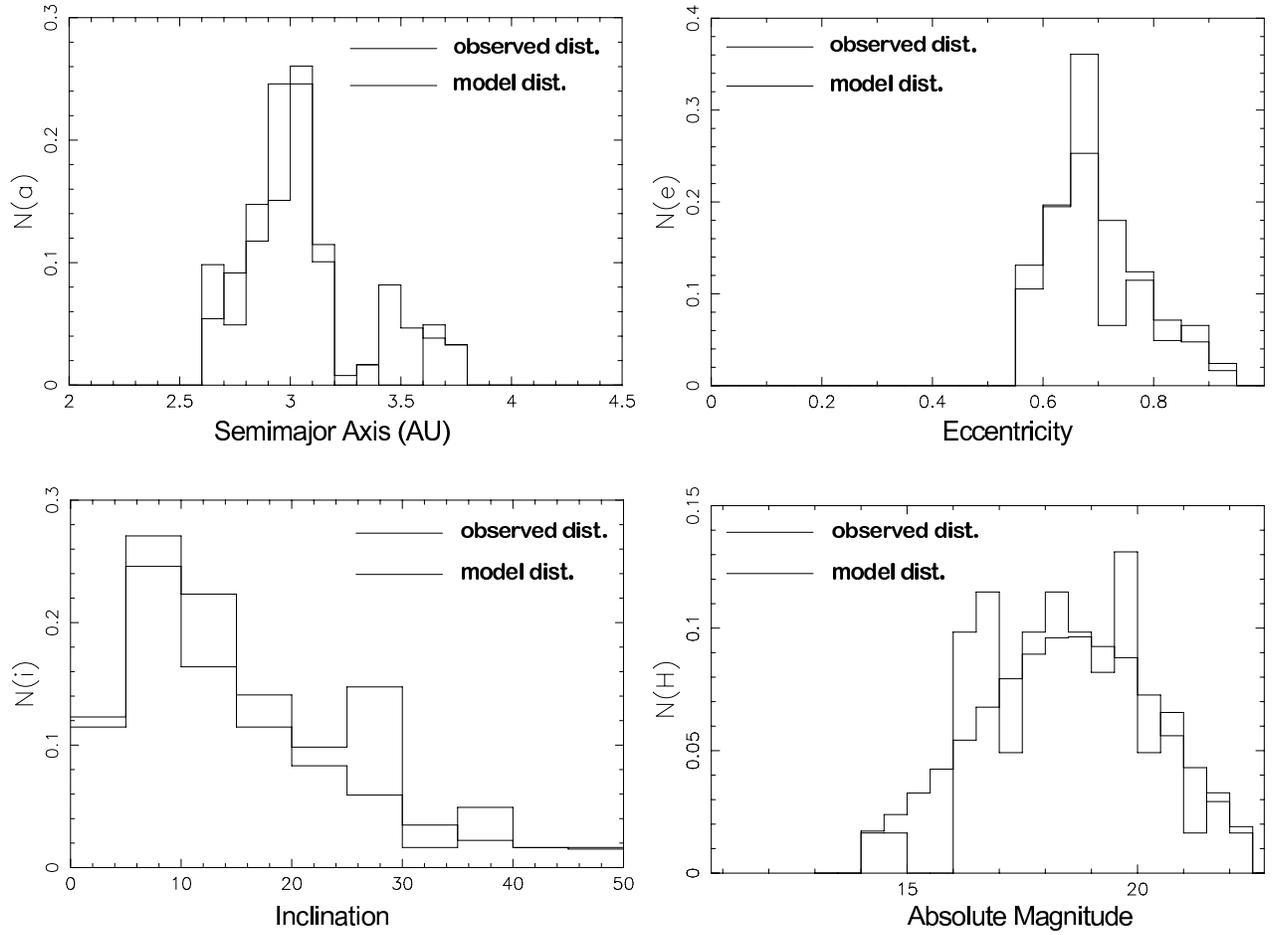}
\caption{Bold histograms: the semimajor axis, eccentricity, inclination, and
  absolute magnitude distributions for dormant JFC candidates among the NEO
  population. Light histograms: the same distributions for our model $B\times
  M_{JFC}$ obtained for the value of $\alpha$ leading to the best quantitative
  match.}
\label{fig.model.data.comparison}
\end{figure}

Our final sample of 61 NEOs are listed in Table
\ref{tab.dormant.JFC.list} while figure \ref{fig.model.data.comparison} shows
the $(a,e,i,H)$ distributions of these objects.  There are many new
unnumbered asteroids listed in the table and their orbital elements
are usually not as accurately known as the numbered asteroids.  This
is of no consequence to our analysis because the binning we will use
in $(a,e,i)$ is much larger than the typical error on unnumbered
asteroid orbital elements and because $P_{JFC}$ usually varies slowly
and smoothly across adjacent bins.

Of course, it is disturbing to base our selection of dormant JFC
candidates on a model. If the \citet{Bot02} model provides a function,
$P_{JFC}(a,e,i)$, that is inaccurate our selected candidates might not
be optimal. To strengthen the validity of our assumptions we will show
(\S\ref{ss.QuantifyingTheStatisticalAgreement}) that our selected
distribution, $m_{JFC}$, is matched by our JFC model distribution,
$B\times M_{JFC}$, at a satisfactory statistical level.  A positive
result implies that, irrespective of how we selected the observed
population, the latter is likely to be dominantly of JFC origin.  In
\S\ref{ss.Dependence} we will also discuss the dependence of the
resulting value of $\alpha$ (the exponent of the $H$-distribution) on
the assumed value of $P_{cutoff}$.

\subsection{Model for NEOs of JFC Provenance}
\label{ss.ModelforNEOsofJFCProvenance}

Following \citet{Bot02}, we used the numerical simulations by
\citet{Lev97} as our model for the orbital distribution of JFCs
. These authors simulated the dynamical evolution of 2,200 test bodies
initially in the trans-Neptunian scattered disk. The orbits of these
bodies were tracked until they entered a major sink or until the
integration time elapsed. Particles reaching $a <2.5$~AU orbits were
cloned 9 times to increase statistics in this zone.

\citet{Bot02} kept track of the amount of time spent by each object
within cells of a 3-dimensional grid in $(a,e,i)$ orbital space,
covering the NEO region ($q=a(1-e)<1.3$~AU) with a resolution of (0.1
AU, 0.05, 5$\arcdeg$).  This ``residence-time distribution''
represents the steady-state relative distribution of JFCs in NEO space
as shown in Fig.~7 of \citet{Bot02}.

Since we selected the dormant JFC candidates as the NEOs with
$P_{JFC}> P_{cutoff}$, for internal consistency we need to restrict
$f_{JFC}(a,e,i)$ to those $(a,e,i)$-bins over which
$P_{JFC}(a,e,i)>P_{cutoff}$ and set the function equal to zero
elsewhere.  Thus, the function $\tilde{f}_{JFC}(a,e,i)$ introduced in
(\ref{e.aeiH-dist}) is re-defined as:
\begin{eqnarray}
f^\prime_{JFC}(a,e,i) & = \tilde{f}_{JFC}(a,e,i)             &{\rm if}\, P_{JFC}(a,e,i) \ge P_{cutoff} \\
f^\prime_{JFC}(a,e,i) & = 0\quad\quad\quad\quad\quad &{\rm if}\, P_{JFC}(a,e,i)  <  P_{cutoff}\ .
\end{eqnarray}
and then $f^\prime_{JFC}$ is normalized to produce $f_{JFC}$.  The
function $f_{JFC}(a,e,i)$ can be regarded as a probability function for the
orbital distribution of JFCs over the restricted $(a,e,i)$ region.

It is important to note that in building $f_{JFC}(a,e,i)$ there is
only a single free parameter: $P_{cutoff}$.  This threshold only
determines the range in the $(a,e,i)$ orbital element space over which
the model is normalized and used.  Furthermore, we will show below
that our result is only weakly dependent on the choice of $P_{cutoff}$
and this is an effect of altering the data rather then altering
$f_{JFC}(a,e,i)$.

We assume that the $H$-distribution is independent of $(a,e,i)$ and
can be expressed in the simple form of equation \ref{e.HFD}.  The
final JFC distribution model $M{JFC}(a,e,i,H)$ is then the product of
$f_{JFC}$ with the $H$-distribution as shown in equation
\ref{e.aeiH-dist}.  The exponent $\alpha$ defining the $H$
distribution is the only free parameter of our model that we will fit
to the observations.  Note that the individual $P_{JFC}$ are
independent of $H$ since they depend only on the orbital elements in
\citet{Bot02}'s residence time distributions.

\subsection{Bias Determination}
\label{s.BiasDetermination}

Many factors determine whether an object will be discovered by a
survey: limiting magnitude, seeing, detector efficiency, weather
conditions, field of view, sky coverage, etc.  The overall probability
that an object with $(a,e,i,H)$ will be discovered is virtually
impossible to calculate analytically except for the most trivial
surveys.  Instead, the bias can be determined through a monte carlo
simulation of the detector's performance that takes into account
important factors determining the observational selection effects
\citep{Jed02}.  The calculation is complicated by the fact that all
the determining factors may vary from field to field even within a
single survey.

Our data sample (\S\ref{ss.TheKnownJFCPopulation}) is comprised of NEOs
discovered by 13 different observatories: Spacewatch, LINEAR, LONEOS, Catalina
Sky Survey, NEAT, CINEOS, AMOS, Palomar, Siding Spring Observatory, Dynic
Astronomical Observatory, Lomnicky Stit, Berne-Zimmerwald, and Haute Provence.
It would be preferable to generate the bias function using detailed
performance criteria from each survey but this information is difficult or
impossible to obtain.  We could have chosen to use only objects discovered by
a single well-characterized survey, but this would have compromised the number
statistics of dormant JFCs within the NEO population available for the study
and decreased the $S/N$ of our result.

Consequently, we chose to follow a different strategy.  \citet{Jed03}
developed a software simulator to imitate, within a single synthetic
survey, the cumulative performance of all real surveys that have
contributed to discovering NEOs.  The simulator incorporated a number
of characteristics typical of asteroid and comet surveys: field
locations, limiting magnitude, minimum detectable rate of motion,
minimum galactic latitude, and detector efficiency.  It passed a
number of tests and detailed comparisons with the performance of
individual well-characterized surveys as discussed in \citet{Jed03}.
Essentially, starting from the \citet{Bot02} model it is been able to
recover the observed orbital and $H$-distribution of real NEOs.  The
simulator has been used by \citet{Jed03} to correctly predict the
fall-off of the NEO discovery rate by the LINEAR survey that has been
observed over the past two years. Moreover, it has been used by
\citet{Lev02} to estimate the fraction of dormant LPCs that surveys
should have detected and, consequently, to determine that most LPCs
disrupt rather than become extinct or dormant.

We have used the \citet{Jed03} simulator to compute the bias function
$B(a,e,i,H)$ in the following manner. We generated a large number of
synthetic objects $G(a,e,i,H)$ and then determined which were
discovered by the simulator, $D(a,e,i,H)$.  The simulated survey bias
is then simply
\begin{equation} 
B(a,e,i,H) = { D(a,e,i,H) \over G(a,e,i,H) }
\end{equation}
with an uncertainty of:
\begin{equation}
\label{eq.sigmaB}
\sigma_{B}(a,e,i,H) = \sqrt{\frac{B(1-B)}{D}}
\end{equation}  

We determined the bias for the candidate dormant JFC nucleii specified
in \S\ref{ss.TheKnownJFCPopulation} ($2.6AU \leq a < 3.8AU$, $.55 \leq
e < .95$, $0^{\circ} \leq i < 55^{\circ}$, $14. \leq H < 22.5$) using
bins of size (0.1 AU, 0.05, 5$\arcdeg$, 0.5) respectively.  There were
a total of 639,360 bins in the calculation.  The synthetic NEO
population, $G(a,e,i,H)$, was the sum of three separately generated
sets of objects:
\begin{itemize}
\item $10^6$ objects distributed evenly among all the bins to ensure
  that $G(a,e,i,H) \neq 0$ for each $a,e,i,H$.  This is important to
  ensure that the bias function is defined in every cell of the model. 
\item $10^6$ objects distributed evenly in $(a,e,i)$, but proportional
  to $10^{0.5H}$ in $H$. This was done in order to generate more
  objects in bins for which the error on the bias determination will
  be large due to small $D$ (see equation \ref{eq.sigmaB}).
  Generating more objects in these bins means that the simulator will
  also find more of them.
\item $10^5$ objects generated in {\it each} of the bins corresponding
  to the 61 NEOs in our sample (see table
  \ref{tab.dormant.JFC.list}).  For example, 100,000 objects were
  generated in the bin occupied by 2004 BZ74 with $3.0 \leq a < 3.1$, 
  $0.85AU \leq e < 0.9AU$, $15.0^\circ \leq i < 20.0^\circ$, 
  $18.5 \leq H < 19$.  This is important to ensure that we
  have a good measure of the bias in each bin incorporated in the
  numerator of the ML method (equation \ref{eq.likelihood-sum}).
\end{itemize}

For all three sets, the orbital angles: mean anomaly, longitude of
node and longitude of perihelion were assumed to be randomly
distributed.  The final synthetic population contained over 8 million
objects that were then ``surveyed'' by the simulator.  The simulation
was run for 2393 synthetic days ($\sim$6.6~y), defined by the time
required by the simulator to discover 694 objects (the number of known
NEOs as of March 5, 2004, the date of the catalogue from which the
list of dormant JFC candidates was extracted) out of the synthetic
population of NEOs in the \citet{Bot02} model.

\begin{figure}
\includegraphics[scale=0.9]{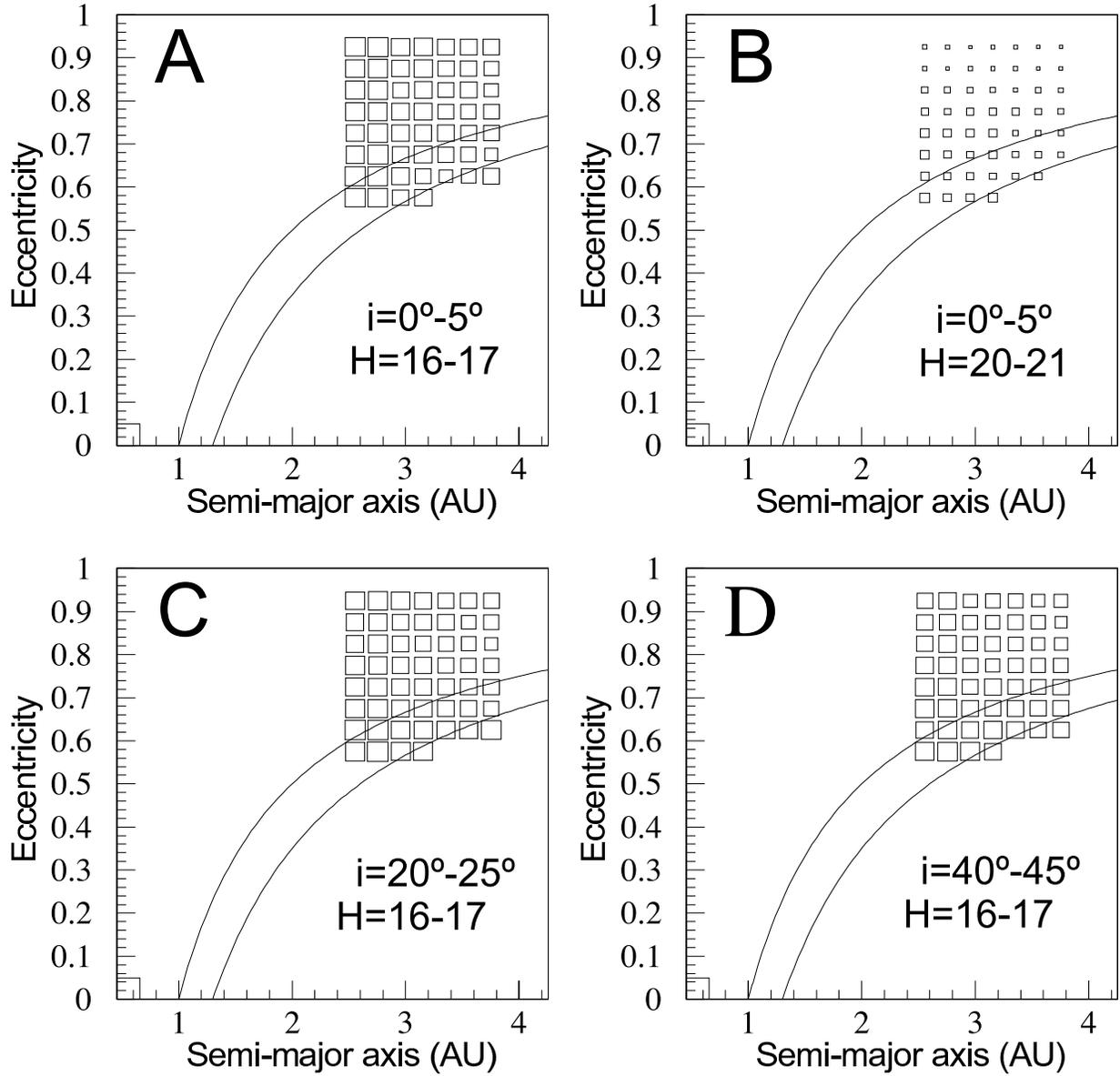}
\caption{Four 2-dimensional $(i,H)$ slices through $B(a,e,i,H)$.  The
ranges of $i$ and $H$ are shown on each of the four figures.
$B(a,e,i,H)$ was calculated only in the orbital element range (see
\S\ref{ss.TheKnownJFCPopulation}) in which the JFCs used in this study
are found.  The size of each box is proportional to the probability
(bias) of discovering an object with orbit elements and absolute
magnitude within the box.  All objects above the lower solid curve are NEOs
(with perihelion $<1.3$ AU) while all objects above the upper solid curve
are Earth crossing (with perihelion $<1.0$ AU).}
\label{fig.jfc.aebias}
\end{figure}

Representative slices through the bias function, $B(a,e,i,H)$, are
provided in figure \ref{fig.jfc.aebias} showing a smooth variation
of the discovery probability as a function of the orbital elements and
$H$.  A comparison of \ref{fig.jfc.aebias}A with
\ref{fig.jfc.aebias}B shows that as the objects become smaller they
become more difficult to detect.  The effect of inclination on the
observational bias is not so strong (compare the progression in
\ref{fig.jfc.aebias}A, \ref{fig.jfc.aebias}C and
\ref{fig.jfc.aebias}D) due to the fact that modern NEO surveys
cover much of the sky to high ecliptic latitudes.  

\subsection{Maximum-Likelihood determination of the slope}
\label{ss.ML}

In this section we lay the mathematical groundwork that is used in the
final calculation of the slope, $\alpha$.  The Maximum-Likelihood (ML)
technique is our method of choice, as it is a powerful tool for
fitting a model to an unbinned data distribution \citep{Lyo86}.  The
ML method determines the parameters of the fit that maximizes the
probability that the model matches the data.

We illustrate the ML method for a function $F(\vec{x},\alpha)$, where
$F$ represents our biased model $B\times M_{JFC}$, the vector
$\vec{x}$ represents the 4-dimensional coordinates $(a,e,i,H)$, and
$\alpha$ is the free parameter.  First, we calculate the normalization
factor $N(\alpha)=\int F \; d\vec{x}$ over the allowed domain of
$\vec{x}$. Thus, $F(\vec{x}^\prime,\alpha^\prime) / N(\alpha^\prime)$
is proportional to the probability that an event with
$\vec{x}=\vec{x}^\prime$ will occur when $\alpha=\alpha^\prime$.  If
there are $n$ events (observations) with $\vec{x}=\vec{x}_i$
($i=1,n$), then the probability of obtaining those $n$ events with the
fit parameter $\alpha$ is proportional to:
\begin{eqnarray}
\label{eq.likelihood-product}
\mathcal L(\alpha) \rm &=& \prod_{i=1}^{n} \; { F(\vec{x}_i,\alpha) \over N(\alpha) }\\ 
                       &=& \prod_{i=1}^{n} \; { F(\vec{x}_i,\alpha) \over \int F(\vec{x},\alpha) \; d\vec{x}}
\end{eqnarray}
Maximizing $\mathcal L$ provides the most probable value of $\alpha$.
The second line of this equation is important because it emphasizes
the fact that the denominator relies on a normalization over the
entire range of $F$ and is a function of the fit parameter ($\alpha$)
while the numerator depends on the $n$ values at the specific
$\vec{x}_i$.

For the purpose of maximimizing equation \ref{eq.likelihood-product}
it is beneficial to take its logarithm to convert the product into a
sum.  Maximizing the logarithm of a function is equivalent to
maximizing the function itself.  The function then becomes:
\begin{equation}
\label{eq.likelihood-sum}
l = \ln(\mathcal L) \rm = \sum_{i=1}^{n} \ln \biggl[ {
    F(\vec{x}_i,\alpha) \over N(\alpha) } \biggr]
\end{equation} 

The value of $\alpha$ when $l$ is at its maximum ($l_{max}$) is the best fit
of the model to the data.  In the ML method, the statistical error on the most
probable result is found by obtaining the values of $\alpha$ at $l_{max} -
1/2$ on both the positive and negative sides of $\alpha$.  The errors are then
$\sigma_{+} = \alpha(l_{max} - 1/2)_{+} - \alpha(l_{max})$ and $\sigma_{-} =
\alpha(l_{max}) - \alpha(l_{max} - 1/2)_{-}$, where the $+$ and $-$ designate
larger and smaller values of $\alpha$ respectively \citep{Lyo86}.

\section{Model results}
\label{s.ModelResults}

Using the procedure described in the previous section and illustrated
in fig. \ref{fig.alpha.vs.P}, we find that the best agreement between model
and observation is achieved with an exponent of the $H$-distribution
for dormant JFCs in NEO space of
\begin{equation}
\label{e.alpha}
\alpha=0.30\pm 0.03(\verb+stat+)\pm 0.05(\verb+sys+).
\end{equation}
The first quoted error is statistical only while the second is an
estimate of the systematic (model-dependent) error introduced by our
technique.  The determination of the systematic error is discussed
briefly in \S\ref{ss.SystematicErrors}.

Before we discuss the implications of this result we need to verify
that our result is meaningful. The fact that we have found a best fit
does not imply that the fit is good. This would be the case, for
instance, if our selected sample of dormant JFC candidates contained
too many asteroids whose orbital distribution cannot be matched with
that of JFCs. Alternative reasons for a bad fit could be that our
orbital distribution model for JFCs is flawed (for instance, the
effects of terrestrial planets or of non-gravitational forces --not
included in the integrations by Levison and Duncan, 1997-- might not
be negligible), or that our evaluation of the biases is unrealistic.
Conversely, if we can show that the fit is good, this would be a
strong indication that all our assumptions for the selection of the
data, construction of the model, and evaluation of the biases, are
reasonable, and that our best fit $\alpha$ is representative of the
real SFD of dormant JFCs.

Figure~\ref{fig.model.data.comparison} shows our best fit model distribution,
$B\times M_{JFC}$, collapsed into 1-dimensional histograms with
respect to $a$, $e$, $i$ and $H$. A visual comparison with the
observed distributions in each parameter suggests that our model
reproduces reality well. However, a qualitative visual agreement is a
necessary but not sufficient condition for a quantitative match
between the model and real 4-dimensional ($a$,$e$,$i$,$H$)
distributions. Testing the actual match between the distributions
requires a numerical analyisis that we develop in the next section.

\subsection{Quantifying the statistical agreement between model and
  observations} 
\label{ss.QuantifyingTheStatisticalAgreement}

In order to quantitatively estimate the statistical agreement between
the model and the observations we have followed the procedure
implemented by \citet{Bot02}. From the best fit model ($B\times
M_{JFC}$, with $M_{JFC}$ obtained with the best fit value of $\alpha$
in the $H$-distribution), we have randomly generated 20,000 synthetic
sets of 61 objects each. Remember that 61 is the number of dormant
JFCs candidates that we selected in
(\S\ref{ss.TheKnownJFCPopulation}), so that each of the synthetic
datasets that we generated has the same number of data points as the
actual data.

By construction, the synthetic datasets `perfectly' match the model.
We then computed the likelihood value, $l$, using
(\ref{eq.likelihood-sum}) for each of the generated 20,000 synthetic
datasets.  The {\it distribution} of $l$ gives the probability that a
likelihood $l^\prime$ is obtained from a dataset that is in perfect
agreement with the model. We found that in 42\% of the synthetic data
sets the likelihood values were {\it lower} than the likelihood
obtained for the actual data. We conclude that our best fit model has
a 42\% probability of being in statistical agreement with the data. In
other words, our model fits the data to within better than 1~$\sigma$
and that we have reproduced the observed distribution of dormant JFC
candidates. In light of this success, all our assumptions in selecting
JFC candidates and in building the model appear justified.

\subsection{Dependence of the best-fit result on the choice of
  $P_{cut}$}
\label{ss.Dependence}

Despite the good results illustrated above it is still legitimate to
ask how our result on the best fit value of $\alpha$ would change if
we chose a more stringent value of $P_{cutoff}$.  As $P_{cutoff}$ is
increased we reduce the contamination of the dormant JFC candidates by
asteroids, but reduce the total number of objects in our sample with a
concommitant increase in the statistical error on the measurement of
the slope.  The dashed line in figure~\ref{fig.alpha.vs.P} shows the
fraction of our original 61 object sample that remains as $P_{cutoff}$
is increased beyond our preferred choice of $0.37$. The solid curve
shows the value of $\alpha\pm\delta\alpha$ determined by the best fit
to the model for the restricted sample.

We see that the calculated value of $\alpha$ is consistent with being
constant and equal to our nominal value over the entire range of
$P_{cutoff}$ values.  On the other hand, there is a small and
systematic decrease in $\alpha$ with $P_{cutoff}$ for
$P_{cutoff}\gtrsim 0.6$.  This might be interpreted in two ways. A
first possibility is that with $P_{cutoff}=0.37$ we have some asteroid
contamination in our data sample (according to the \citet{Bot02} model
about 1/3 of our selected objects should be asteroids). So, the value
of $\alpha$ that we determine is a weighted average between the slope
characterising NEOs of asteroidal origin ($\alpha=0.35$ according to
\citet{Bot02}) and the actual $\alpha$ of dormant JFCs. By increasing
$P_{cutoff}$ we reduce the asteroid contamination and the genuine
value of $\alpha$ for the JFCs is exposed. The second possibility is
that $\alpha$ changes simply because we are fitting a smaller dataset,
but the results are still statistically consistent with $\alpha=0.30$
as shown in figure~\ref{fig.alpha.vs.P}. The fact that our choice of
$P_{cutoff}$ maximises the signal/noise ratio of the JFC sample
(\S\ref{ss.TheKnownJFCPopulation}), and that the corresponding data
set is well matched by our model, makes us confident that $\alpha\sim
0.30$ is a measure of the real slope of the $H$-distribution of
dormant JFCs.

\subsection{Systematic errors}
\label{ss.SystematicErrors}

\begin{figure}
\includegraphics[angle=0,scale=.7]{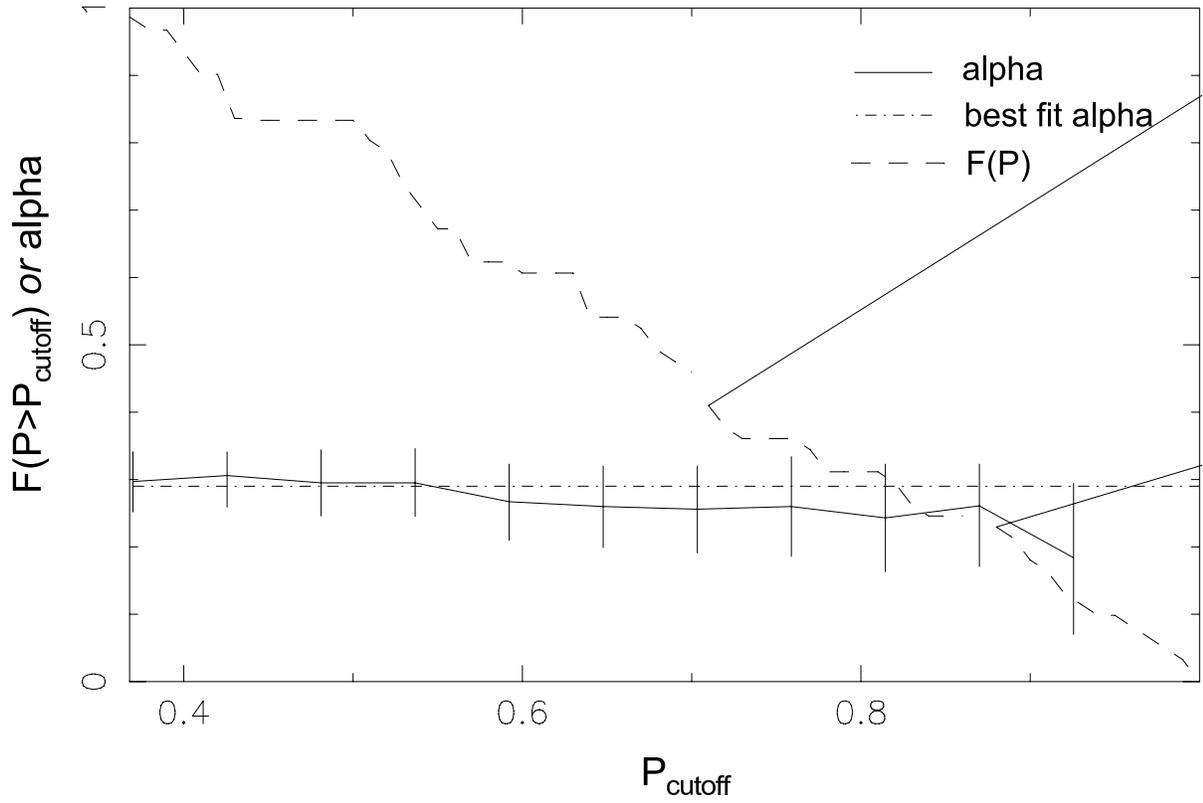}
\caption{Dashed line: the fraction of our original 61 NEO sample that
  remains selected if $P_{cutoff}$ is increased beyond our preferred
  value of 0.37.  Solid line: the value of $\alpha$ obtained by best
  fit of the model to the restricted NEO sample of dormant JFC
  candidates as a function of $P_{cutoff}$. The horizontal dash-dotted
  line indicates our preferred value of $\alpha=0.30$.}
\label{fig.alpha.vs.P}
\end{figure}

The error bars on the best fit values of $\alpha$, illustrated in
figure~\ref{fig.alpha.vs.P}, represent the statistical errors given by the ML
fitting method (\S\ref{ss.ML}).  However, the technique itself is
model-dependent and introduces systematic errors into our $\alpha$
determination.  As always, the systematic error should be quantified
wherever possible.

There are a number of model-dependent sources of error in our slope
determination.  Our use of the nominal \citet{Bot02} model for the JFC
probability as a function of ($a$,$e$,$i$) could be a source of
systematic error.  If the actual dynamical distribution or
contributions of the source populations going into that model is
different from the assumption it would affect our slope determination.
We consider it unlikely that the dynamical distribution of the source
populations could be much different from that used in the
\citet{Bot02} analysis because it used high-statistics and a dynamical
integrator of good pedigree.  On the other hand, the contribution from
each of the 5 source regions (or even the inclusion of other possible
sources) was determined from a fit to the model for a small number of
data points.  The relative contribution from each source could be
different from the nominal model.  Recent, as-yet-unpublished work
(Bottke, personal communication) suggests that the inclusion of
10$\times$ more data using the results of the LINEAR survey have
little effect on the fraction of the NEO population deriving from each
source population.

To estimate this systematic effect, let $f_{JFC}$ represent the
fraction of the NEO population from the JFC source region and
$p_{JFC}(a,e,i,H)$ be the dynamical probability that a JFC will appear
in the bin near $(a,e,i,H)$.  Similarly, let $f_x$ represent the
fraction of the NEO population from all other sources and
$p_x(a,e,i,H)$ be the dynamical probability that an object from any
other source will appear in the same bin.  Then the probability that
an object is a JFC in that bin is simply
\begin{equation}
P_{JFC} = { f_{JFC} \; p_{JFC} \over f_{JFC} \; p_{JFC} + f_x \; p_x }.
\end{equation}
Recognizing that $f_{JFC}=1-f_x$ and differentiating with respect to
the JFC fraction ($f_{JFC}$) we find that
\begin{eqnarray}
dP_{JFC} &=&       \biggl[ p_{JFC} - p_{JFC}^2 \; \biggl( 1 - { p_x \over
    p_{JFC} } \biggr) \biggr] \; { df_{JFC} \over f_{JFC} } \\
dP_{JFC} &\approx& p_{JFC} \; ( 1 - p_{JFC} ) \; { df_{JFC} \over f_{JFC} }
\end{eqnarray}
where we have made the approximation that $p_x/p_{JFC}\ll 1$ since we
are expressly working with a region in $(a,e,i)$-space that has a
high-probability of containing objects of JFC provenance.
\citet{Bot02} calculated that $f_{JFC}=0.06\pm 0.04$ and we used
$P_{JFC}=0.37$ so that we expect the effect of an inaccurate
determination for the fraction of the NEO population that are JFCs to
be on the order of $dP_{JFC}\sim 0.15$.  Figure~\ref{fig.alpha.vs.P} shows
that the calculated value for the slope is relatively insensitive to
changes in $P_{JFC}$ of this magnitude so that we consider this to be
a negligible source of systematic error.

Figure~\ref{fig.alpha.vs.P} further explores the systematic error
introduced by our selection of $P_{cutoff}$.  In that figure we vary
that value over the range from 0.37 to 0.7 and obtain best fit values
for $\alpha$ ranging from 0.310 to 0.256.  We associate half this
difference, or 0.027, with the systematic error introduced due to the
data selection process.

Another source of systematic error is the bias estimate.  As we
explained in \S\ref{s.BiasDetermination}, to compute the bias function
we performed a survey simulation for a large population of realistic
synthetic NEOs \citep{Bot02}. The parameters of the survey simulator
were empirically `tuned' to provide a distribution of `discovered'
objects that mimicked the known distribution \citep{Jed03}.
Nevertheless, some uncertainty remains in the survey simulator's
parameters. The most important one is the limiting magnitude of the
simulated survey. In \S\ref{s.BiasDetermination} we used
$V_{lim}=20.5$ but a choice of $V_{lim}$ 0.5 magnitudes fainter or
brighter would give a similarly good reconstruction of the observed
NEO population. Thus, we computed two additional bias functions by
running the simulator with $V_{lim}=20.0$ and $V_{lim}=21.0$. The
number of virtual days over which the simulation was run was also
changed so that the total number of discovered synthetic NEOs remained
equal to the actual number of NEOs found by real surveys. Then, using
these new bias functions, we repeated the fitting procedure detailed
in this paper and found that the best fit value of $\alpha$ 
changed by $\pm 0.04$ with respect to our nominal value of 0.30.


Thus, we estimate that the magnitude of the systematic error on the
best fit slope for the $H$-distribution is the quadratic sum of the
two values given above or $0.05$ as stated in equation \ref{e.alpha}.

\section{Discussion}
\label{s.Discussion}

Using our best fit model, we can estimate the total number of dormant
JFCs. For this purpose, we first need to evaluate the weighted mean
bias $\bar{B}$, defined as:
\begin{equation}
\bar{B}=\sum_{(a,e,i)} \; \sum_{14<H<22.5} B(a,e,i,H) \; M_{JFC}(a,e,i,H)
\end{equation}
where the first sum is computed over all the $(a,e,i)$ cells for which
$P_{JFC}\ge P_{cutoff}$, $M_{JFC}(a,e,i,H)$ is our best fit
normalized model (where $\sum_{(a,e,i)} \; \sum_{14<H<22.5} \;
M_{JFC}(a,e,i,H) = 1$), and we only sum over bins for which
$B(a,e,i,H)\ne 0$.  We find $\bar{B}=8.22\times 10^{-3}$.

Thus, the 61 dormant JFCs discovered over the same $(a,e,i,H)$ range
imply a total population of $\sim 1,400$ objects with $H<22.5$. Given
our best fit value of $\alpha$, the population of dormant JFCs with
$H<18$ is $1,400\times 10^{[-0.30\times (22.5-18)]}\sim 63$. However,
this is the number within the restricted region where
$P_{JFC}(a,e,i)>0.37$ rather than the total number of JFCs in the
entire NEO region.  From the orbital distribution model that we have
adopted (\S\ref{ss.ModelforNEOsofJFCProvenance}), we compute that the
total number of JFCs with $a<7.4$~AU and $q<1.3$~AU (i.e. JFCs in the
NEO region) should be larger by a factor of $\sim$1.2, implying $\sim
75$ objects with $H<18$.  There are about 25 known objects in this
size range in our data sample implying a completion rate of about 1/3
for dormant JFCs with $H<18$.  This may be compared to the current
completion of about 70\% for all NEOs with $H<18$.  The reduced
completion rate for the dormant JFC NEOs is easily understood as an
observational selection effect due to their large eccentricity and
correspondingly larger heliocentric distance.

This number is in good agreement with that of $61\pm 43$ determined by
\citet{Bot02} and we stress that our result is independent of their
calculation. The only thing that the present work has in common with
the earlier work is the orbital distribution model for JFCs. The bias
function and the data used to fit the model are different. The use of
$P_{JFC}$ from \citet{Bot02} in the selection of our data is only
incidental as discussed in \S\ref{ss.ModelforNEOsofJFCProvenance}.
Thus, the agreement between the two results strengthens both
models. 


\citet{Fer99} claim that there are $30_{-5}^{+10}$ active JFCs with
$H<18$ and $q<1.3$~AU while \citet{Lev97} suggest that for each active
comet there should be 3.5 faded comets (ranging from a minimum of 2.0
to a maximum of 6.7). Thus, assuming that all faded comets are
dormant, we estimate that the total number of dormant JFCs with $H<18$
and $q<1.3$~AU is $30 \times 3.5 = 105$, with a possible range from 50
to 270.  Our estimate of the actual number of dormant JFCs falls at
the lower end of this range but not too far from the nominal value.
We conclude, as in \citet{Bot02}, that a substantial fraction of JFCs
become dormant when they fade.  Unlike in the LPC and HTC cases (for
which only $\sim 1$\% of faded comets seem to survive in a dormant
state according to \citet{Lev02}), disintegration is likely not the
explanation for the disappearance of JFCs or their final fate.

In the previous section we showed that the slope of the
$H$-distribution of dormant JFCs is $\alpha= 0.30\pm 0.06$ (combined
statistical and systematic error).  This value is in agreement with
that determined by \citet{Lev02} for a sample of 9 dormant HTCs
($\alpha_{HTC}=0.23\pm 0.04$) despite the different fading behaviors
of the two types of comets as described in the last paragraph.

As we stated in the introduction, it is instructive to compare the
size distribution of active and dormant comets.  Unfortunately,
measuring the SFD of active comets or, equivalently, the $H$-magnitude
distribution of their {\it nuclei}, is a daunting task.  Since comets
are small they are easiest to observe when closest to Earth, but at
this heliocentric distance they develop comae that mask their nuclei
and can increase their brightnesses by up to 10 magnitudes
\citep{Fer99}.  (See \citet{Lam04} for a recent review and compilation
of cometary nuclei.)

The observation of an active comet leads, with some assumption of the
dependence of its activity on the heliocentric distance, to an
estimate of the comet's {\it total absolute magnitude}, often denoted
$H_T$ or $H_{10}$. This is a measure of the intrinsic brightness of
the combined nucleus and coma. Some studies have attempted to
physically model the coma in order to subtract its contribution to the
overall brightness profile and reveal the bare nucleus
\citep[discussed in detail in][]{Tan00}.  This method involves
modelling light scattered by dust grains and gas (due to sublimation
from the surface) ejected from the nucleus.  The modelling process is
difficult and different coma profiles yield various estimates of the
nucleus' brightness. \citet{Fer99} found that for very active comets
$H_T$ scales as $0.75 H_N$, while for low-activity comets it scales as
$1.5 H_N$. $H_N$ is the absolute magnitude of the bare nucleus, or the
absolute magnitude that the comet would have if it had no activity.
It can be identified as the absolute magnitude $H$ used throughout
this work.

\citet{Hug02} reports that the $H_T$-distribution of the active bright
JFCs ($H_T<6.6$) is $\alpha_T$(JFC)=0.36. (This value is essentially
the same as that determined by the same author for active, bright LPCs
\citet{Hug01}.)  If one accepts the scaling between $H_T$ and $H$ for
very active comets reported above, the $H$-distribution of active JFCs
would have an exponent $\alpha_N$(JFC)=0.24.  Fernandez and Morbidelli
(work in progress) found that the $H_T$ distribution for faint active
JFCs has $\alpha_T$(JFC)=0.2. Using the scaling between $H_T$ and $H$
for low activity comets reported above, this gives
$\alpha_N$(JFC)=0.27.  Thus, once again, there is good agreement
between the results presented here and other recent work.

Another approach to determining the nuclear SFD involves measuring
cometary magnitudes at large heliocentric distances where the comets
are presumably inactive.  \citep[\eg][]{Mee04, Low03}.  There are
problems with this approach as well since cometary nuclei are very
faint at large heliocentric distances and magnitude measurements are
affected by sky brightness and seeing.  It is also known that some
comets remain active up to many AU from the sun, and that this
residual activity is often difficult to detect due to the small angle
subtended by a distant comet on the sky.  

One stab at the problem involves comparing a comet's profile with a
stellar PSF.  If the comet is extended compared to the point source
then it is considered to have a coma which can then be modeled and
subtracted away \citep{Lic00}.  This method will not be effective for
comets so far away that they are completely contained within a pixel -
\eg\ \ a comet located at 10AU with a 7200 km diameter coma ($\sim$8
orders of magnitude larger in volume than the typical nucleus)
subtends only 1\arcsec.  Thus, even very large comae that contribute
dramatically to the apparent brightness of a distant comet can go
unnoticed.

Using the coma modelling and subtraction approach, \citet{Fer99} found
that the $H$-distribution of JFCs has $\alpha=0.53\pm0.05$ over a very
narrow magnitude range corresponding to nuclear radii of 2--4~km.
\citet{Wei03} found $\alpha=0.32\pm0.01$ in the 1-10~km range while
\citet{Mee04} found $\alpha=0.290\pm0.010$ in the same range and
$\alpha=0.382\pm0.012$ in the sub-range from 2-5~km.  \citet{Lam04}
summarized all these results in their Table 6 where they also report
an updated value for the \citet{Wei03} paper by Weissman of
$\alpha=0.36\pm0.01$.  Furthermore, \citet{Lam04} performed their own
analysis of the cumulative data on cometary nucleii and found
$\alpha=0.38\pm0.06$ for nucleii $>$3.2~km in diameter.  These may be
compared to our result of $\alpha=0.30\pm0.06$ for dormant cometary
nucleii diameters in the range 0.25-15~km (assuming a 0.04 albedo
typical of cometary nucleii, \citet{Fer01}).  It is interesting to
note that the two studies with largest number of objects have the
largest reported errors on their slopes.

It should be pointed out that the bulk of the JFC candidates on which
we have based our study have $16<H<21$ corresponding to radii between
0.1~km and 2~km.  Thus, our $\alpha$ applies to a size range that is
not really addressed by the studies of active comet SFDs mentioned
above, except in the preliminary Fernandez and Morbidelli work.
Despite all the uncertainties, the convergence and agreement of all
the slope measurements using different techniques (except for the
single result of \citet{Fer99}) gives an indication that the SFD of
active comets is shallow.

Keeping all these caveats in mind, the similarity between the SFD of
active and dormant comets suggests that in the fading process the
probability of becoming dormant versus disintegrating is roughly size
independent. This seems to be true both when comets in the majority
become dormant (as in the JFC case) and when they most likely disrupt
(as in the LPC/HTC case).

This conclusion can guide us to an understanding of why the SFD of
{\it active} comets is shallow. In essence, there are two possible
interpretations. Either (i) the SFD of comets in the reservoirs
(scattered disk, Oort cloud) is as shallow as observed for active
comets, or (ii) the SFD of active comets is shallower than that in the
parent reservoirs as a consequence of a size-dependent fading
probability.  The fact that the SFD of dormant comets is similar to
that of active comets tends to support (i) against (ii). The reason is
simple.  Imagine the case that comets have a size-dependent fading
probability and that all faded comets are dormant. Then, the SFD of
active comets would appear shallower than the parent SFD, but the SFD
of dormant comets would appear {\it steeper}.  The same happens in the
case where a size-independent fraction of faded comets become dormant.
Thus, scenario (ii) requires that the probability of remaining active
and the probability of becoming dormant have the same size-dependence
in order that both active and dormant populations have the same
SFDs. There is no physical reason for this to be true so if it was it
would be a striking coincidence. Thus we prefer scenario (i) as an
explanation for the shallow cometary SFD.

The difficulty is that scenario (i) raises the problem of explaining
why the populations in the comet reservoirs are shallow.  Both the
scattered disk and the Oort cloud preserve a fraction of the
population of planetesimals initially in the protoplanetary disk
through which the giant planets migrated \citep{Don05}. Thus, although
the current collisional activity inside the comet reservoirs is
minimal, originally the comets had to belong to a massive small body
population where the collisional evolution had to be intense (at least
during the initial phase of transport towards the scattered disk/Oort
cloud; see \citet{Ste01} and \citet{Cha03}), and thus their SFD had to
be close to that of a population at collisional equilibrium. It is now
known that, because the impact strength of the planetesimals is size
dependent and has a minimum at about 100~m in radius \citep{Ben99},
the equilibrium size distribution is very shallow in the range 100~m
-- 5~km.  In fact, according to the SDSS survey \citep{Ive01} the main
asteroid belt (the best example we have of a small body reservoir in
collisional equilibrium), has an $H$-distribution with $\alpha\sim
0.26$ in the range between 300~m and 5~km in diameter. Thus, it is
plausible that the SFD of active comets is representative of the SFD
in the scattered disk and in the Oort cloud, which in turn are both a
fossil remnant of the shallow SFD in the original protoplanetary disk.
The Kuiper belt might have also a similarly shallow SFD in the same
size range \citep{Pan05}.

If this is true, the SFD of comets should become steep again below 100
meters in size. We do not have any direct observational evidence of
the populations of active and dormant JFCs in favor or against this
prediction. These bodies are simply too faint to detect. However, the
paucity of small primary craters on Europa \citep{Bie05}, suggests
that the SFD of comets remains shallow below 100~m in size. A favored
explanation is that objects below this size threshold cannot stand the
thermal shocks suffered as they approach within 5--10~AU of the Sun,
and therefore disintegrate before they evolve into Jupiter crossing
orbits. If this is true, we need to modify the scenario above as
follows.

The cometary activity of the large Centaur Chiron suggests that as
comet precursors penetrate into the Centaur region (non-Jupiter
crossing orbits with $5.2<a<30$~AU) they experience thermal loads that
induce sublimation of volatile ices (\eg\ N$_2$, CO, CH$_4$, H$_2$CO,
NH$_3$, CO$_2$, see \citet{Del82}).  In order to explain our results
we propose that there exists a size-dependent disintegration of the
precursor objects before they become `comets'.  We will not speculate
deeply on the physical mechanism of disruption but point out that the
repeated and increasing thermal pulses suffered by the objects as they
dynamically evolve closer to the Sun might induce internal mechanical
stresses.  The smaller the initial object, the more likely it may be
to disrupt under the thermal-induced stresses that it is experiencing
for the first time.  Consequently, the population of bodies smaller
than 100 meters is annihilated and the SFD of the larger survivors
might become somewhat shallower than the SFD of the parent population.
The bodies that have passed this initial decimation are strong enough
to be able to penetrate closer to the Sun, develop cometary activity
for some perihelion passages, and finally form a crust of refractory
material that eventually makes them dormant.

This scenario is different from the scenario (ii) that we
have rejected above.  Scenario (ii) was invoking a size-dependent
fading probability.  This scenario requires a size-dependent
disruption probability before the beginning of classical cometary
activity. Then, for the comets that pass this first selection and
develop cometary activity, it invokes a size-independent fading to the
dormant state as given in scenario (i).  This new scenario
(i$^\prime$) would explain the shallow SFD of active JFCs, the equally
shallow SFD of dormant JFCs and the paucity of small primary craters
on Jupiter's satellites.  It is a little bit more problematic to
explain why the SFDs of the active LPCs and the dormant HTCs are also
similar.

With the currently available data, it is difficult to discriminate
between scenarios (i) and (i$^\prime$).  To break the degeneracy
between the two requires additional information in the form of (a) the
detection of smaller active/dormant comets in order to extend the SFD,
and (b) an in-situ detailed analysis of the crater SFD on the distant
icy satellites of all giant planets and Kuiper belt objects.

\section{Conclusion}
We have identified a set of 61 NEOs that are likely to be the dormant
nucleii of JFCs and modelled their observational selection effects in
order to determine the slope of their actual absolute-magnitude
frequency distribution: $\alpha=0.30\pm 0.03(\verb+stat+)\pm
0.05(\verb+sys+)$.  The total number of dormant JFCs with $H<18$ in
the NEO region is $\sim 75$ and the completion rate in the known data
sample for objects in this size range is about 30\%.  The slope for
the HFD as determined by our novel method is consistent with most
other recent measures of the same quantity that used different
techniques \citep[\eg\ ][]{Fer99,Wei03,Mee04}.  Our results push the
HFD to a smaller size range than have been published and are also
applicable over a wider range than previous work.  The estimate for
the total number of dormant JFCs is consistent with other estimates
\citep{Bot02}.

Our results suggest a physical and dynamical evolutionary scenario
as follows:

\begin{itemize}
\item The SFD of comets in their reservoirs (scattered disk, Oort
cloud) is shallow and a remnant of their equilibrium SFD within the
proto-planetary disk before being scattered.

{\it or}

\item The SFD of comets in their reservoirs is somewhat shallow and as
the comet precursors penetrate into the giant planet region
(non-Jupiter crossing orbits with $5.2<a<30$~AU) they suffer strong
thermal stresses that induce a size-dependent disintegration
probability before they become `comets'. Objects smaller than 100~m
diameter are effectively annihilated and the SFD of the survivors will
be shallower than the SFD of the parent population.

{\it then}

\item The objects penetrate closer to the Sun, develop
cometary activity for some perihelion passages, and finally form a
crust of refractory material that eventually makes them dormant.  All
these processes proceed in a size-independent manner.
\end{itemize}

These scenarios explain the observed shallow SFD of active JFCs, the
equally shallow SFD of dormant JFCs (this work and \eg\ 
\citet{Fer99,Wei03,Mee04}) and the paucity of small primary craters on
Jupiter's satellites \citep{Bie05}.

Future studies involving the detection and characterization of smaller
active/dormant comets, and in-situ analysis of the crater SFD on the
distant icy satellites of all giant planets and Kuiper belt objects
will provide the litmus test for deciding if this scenario is realistic.

\section{Acknowledgements}
We thank Bill Bottke, Yanga Fern{\' a}ndez, Hal Levison and Nalin
Samarasinha for helpful discussions interpreting the results of this
study.  Paul Weigert and Philippe Lamy provided constructive reviews.


\clearpage
\begin{deluxetable}{lccccc}
\tablewidth{0pt}
\tablecaption{Dormant JFC candidates selected from known NEOs}
\tablehead{
\colhead{Designation}  & 
\colhead{a} &
\colhead{e} &  
\colhead{i} & 
\colhead{H} &
\colhead{$P_{JFC}$}
}
\startdata
   2004 BZ74 &   3.05&  0.893 &  16.6 &  18.7 &  0.98\\
   2003 YS1  &   3.10&  0.847 &  25.1 &  19.7 &  0.42\\
   2003 WY25 &   3.08&  0.675 &   5.9 &  21.1 &  0.88\\
   2003 WR25 &   3.35&  0.710 &   9.0 &  19.6 &  0.94\\
   2003 UO12 &   2.74&  0.700 &  45.1 &  15.4 &  0.54\\
   2003 HP32 &   2.69&  0.779 &   3.4 &  19.7 &  0.51\\
   2002 XE84 &   2.82&  0.663 &  28.9 &  20.7 &  0.44\\
   2002 UO3  &   2.96&  0.802 &  24.1 &  17.8 &  0.91\\
   2002 MT3  &   2.81&  0.690 &   6.5 &  19.9 &  0.64\\
   2002 KG4  &   2.94&  0.663 &  27.6 &  20.9 &  0.78\\
   2002 JB9  &   2.72&  0.785 &  46.7 &  16.0 &  0.55\\
   2002 GZ8  &   2.79&  0.653 &   5.3 &  18.4 &  0.39\\
   2002 GJ8  &   2.96&  0.828 &   5.3 &  19.3 &  0.98\\
   2002 FC   &   2.83&  0.661 &   6.8 &  19.0 &  0.64\\
   2002 EX12 &   2.60&  0.767 &  11.3 &  16.1 &  0.55\\
   2002 CX58 &   2.80&  0.659 &   2.5 &  22.2 &  0.54\\
   2001 YK4  &   2.65&  0.778 &   4.6 &  18.5 &  0.51\\
   2001 XP1  &   2.90&  0.751 &  39.3 &  17.8 &  0.78\\
   2001 QN142&   3.09&  0.686 &  10.2 &  21.8 &  0.90\\
   2001 ME1  &   2.65&  0.865 &   5.8 &  16.9 &  0.41\\
   2000 YG29 &   3.17&  0.695 &  18.9 &  18.8 &  0.78\\
   2000 WL10 &   3.16&  0.714 &  10.2 &  18.0 &  0.93\\
   2000 PG3  &   2.83&  0.858 &  20.5 &  16.2 &  0.93\\
   2000 KE41 &   3.00&  0.865 &  50.4 &  17.4 &  0.82\\
   2000 DN1  &   2.88&  0.670 &   7.8 &  19.7 &  0.64\\
   1999 UZ5  &   2.64&  0.799 &  10.4 &  21.8 &  0.55\\
   1999 RD32 &   2.64&  0.771 &   6.8 &  16.6 &  0.53\\
   1998 SY14 &   2.85&  0.665 &   3.5 &  20.6 &  0.54\\
   1998 FR11 &   2.79&  0.713 &   6.6 &  16.4 &  0.65\\
   1984 QY1  &   2.97&  0.917 &  15.5 &  14.0 &  0.96\\
   1986 JK   &   2.85&  0.665 &   2.0 &  18.3 &  0.54\\
   2003 SD201&   3.03&  0.640 &  20.9 &  17.9 &  0.71\\
   2003 RM   &   2.91&  0.604 &  10.9 &  20.1 &  0.43\\
   2003 LO6  &   2.91&  0.576 &  34.6 &  16.8 &  0.69\\
   2003 AC1  &   3.14&  0.653 &  23.5 &  20.7 &  0.57\\
   2002 WW17 &   3.02&  0.654 &  18.4 &  17.6 &  0.83\\
   2002 VT94 &   3.09&  0.587 &  25.1 &  19.7 &  0.90\\
   2002 UN   &   3.01&  0.609 &  26.2 &  17.3 &  0.94\\
   2002 RC118&   2.95&  0.565 &  28.0 &  16.8 &  0.72\\
   2002 RN38 &   3.80&  0.675 &   3.8 &  17.3 &  1.00\\
   2002 AO7  &   2.94&  0.626 &  14.9 &  18.2 &  0.43\\
   2002 AR4  &   3.00&  0.622 &   8.3 &  20.0 &  0.80\\
   2001 XQ   &   3.64&  0.713 &  29.0 &  19.5 &  1.00\\
   2001 UU92 &   3.17&  0.669 &   5.4 &  20.1 &  0.68\\
   2001 TB45 &   3.00&  0.576 &  25.1 &  19.0 &  0.72\\
   2001 SK169&   3.01&  0.568 &  20.2 &  17.6 &  0.67\\
   2001 AO2  &   3.07&  0.609 &  19.9 &  18.3 &  0.60\\
   2000 LF6  &   2.91&  0.611 &  14.8 &  19.9 &  0.43\\
   2000 EB107&   3.03&  0.585 &  25.3 &  16.9 &  0.90\\
   1999 VX15 &   3.01&  0.600 &  12.3 &  18.8 &  0.72\\
   1999 LT1  &   2.98&  0.657 &  42.6 &  17.6 &  0.74\\
   1999 DB2  &   3.00&  0.620 &  11.6 &  19.1 &  0.72\\
   1998 SE35 &   3.01&  0.593 &  14.8 &  19.2 &  0.41\\
   1998 HN3  &   3.12&  0.618 &   9.2 &  18.4 &  0.55\\
   1998 GL10 &   3.18&  0.668 &   8.7 &  18.5 &  0.68\\
   1997 SE5  &   3.73&  0.666 &   2.6 &  14.8 &  1.00\\
   1992 UB   &   3.07&  0.581 &  15.9 &  16.3 &  0.42\\
   1982 YA   &   3.66&  0.700 &  35.3 &  16.5 &  0.97\\
   1982 YA   &   3.63&  0.697 &  35.0 &  18.1 &  0.83\\
   1998 MX5  &   2.98&  0.612 &   9.7 &  18.5 &  0.58\\
   1994 LW   &   3.19&  0.617 &  22.4 &  16.9 &  0.71
\enddata
\label{tab.dormant.JFC.list}
\end{deluxetable}


\clearpage

\clearpage

\clearpage

\clearpage

\clearpage

\clearpage

\end{document}